\documentclass[12pt]{article}  

\usepackage{psfig}
\begin{document}
\title{A codebook generation algorithm for document image compression}

\author{  Qin Zhang \and John M. Danskin \and Neal E. Young}
\date{    6211 Sudikoff Laboratory\\
         Department of Computer Science\\
         Dartmouth College\\
         Hanover, NH 03755, USA\\
         \{zhangq, jmd, ney\}@cs.dartmouth.edu
         }

\maketitle
\thispagestyle{empty}
\begin{abstract}
\small
  Pattern-matching based document compression systems rely on finding a small
  set of patterns that can be used to represent all of the ink in the document.
  Finding an optimal set of patterns is NP-hard; previous compression schemes
  have resorted to heuristics.  We extend the cross-entropy approach, used
  previously for measuring pattern similarity, to this problem.  Using this
  approach we reduce the problem to the fixed-cost $k$-median problem, for
  which we present a new algorithm with a good provable performance guarantee.
  We test our new algorithm in place of the previous heuristics (First Fit,
  with and without generalized Lloyd's (k-means) postprocessing steps).  The
  new algorithm generates a better codebook, resulting in an overall improvement
  in compression performance of almost 17\%.
\end{abstract}

\section{INTRODUCTION}
For scanned text, pattern matching based compression
achieves the best compression ratios currently known:
roughly four times that of 2D-FAX coding \cite{witten:94}.
Pattern matching based compression has been studied
by several research groups \cite{ascher:74,holt:86,mohiuddin:84,witten:94}.
This kind of compression involves the following steps:
\begin{description}
\item[\bf Extract] from the image of the scanned document a sequence of
  {\em glyphs}. Each glyph typically represents one connected blob of ink
  occuring somewhere in the document, represented as a positioned bitmap.
\item[\bf Partition] the set of glyphs into equivalence classes of
  glyphs with similar bitmaps; for each class, compute a representative
  {\em pattern} --- a bitmap representing the bitmaps of all glyphs in the
  class.  This set of canonical patterns is the {\em codebook}.
\item[\bf Transmit] the codebook and, for each glyph in the sequence of
  glyphs representing the document, the glyph's position and the index of its
  representative pattern.  For {\em lossy\/} compression, the process stops
  here.  For {\em lossless\/} compression, for each glyph,
  also transmit the difference between the glyph and its representative,
  so that the glyph's bitmap can be exactly reconstructed.
\end{description}
For effective compression, the transmission step
requires additional compression techniques 
to economically represent the codebook,
indices, and position information.
For this work,
the codebook is compressed using Moffatt's two-level context-based method
(following \cite{witten:94}).
The glyph indices are coded using PPMC.
The glyph positions are coded using structure-based position coding
developed in~\cite{zhang:ep96}.

Several recent papers have focussed on how to measure the {\em distance\/}
between pairs of bitmaps \cite{inglis:94,zhang:96} via cross-entropy
measures.
These measures approximate the number of bits needed to encode one bitmap
given the other under various models.
In this paper we use the measure proposed in \cite{zhang:96}.

The main contribution of this paper
is a new method for the {\em partitioning} step.
Recent works have used relatively ad-hoc heuristics for this step,
with the most succesful bearing a resemblance to
Lloyd's algorithm for vector quantizer design \cite{gersho}.

The conceptual basis of the new method
is an extension of a particular cross-entropy model,
proposed and approximated in \cite{zhang:96}, for measuring distances between bitmaps.
Here is a summary of this model.
Each bitmap is assumed to have been generated by scanning an ideal character
(a blob of ink) at some random offset.
Thus, any ideal character induces a probability distribution on the bitmaps.
Conversely, any bitmap induces a probability distribution on the ideal
characters --- the uniform distribution on those ideal characters consistent
with the bitmap.  The {\em patterns\/} in this model are these distributions.
Each such pattern has a succint representation as a bitmap.
The {\em distance\/} between a pattern and a glyph is the conditional entropy
of the glyph given the distribution on ideal characters associated with the
pattern.  

To extend this model to the partitioning problem,
we make the useful but perhaps unwarranted assumption
that the a-priori distribution of patterns is uniform.
With this assumption, given the set of glyphs in the document,
each set of possible patterns has a conditional probability.
We call the set of patterns with the maximum conditional probability
the {\em minimum-entropy\/} codebook.
We define the {\em cost\/} of a pattern to be the number of bits
needed to represent it,
the {\em cost\/} of a codebook to be the sum of the costs of its patterns,
and the {\em distortion\/} of a codebook to be the number of bits
needed to represent all the glyphs in the document given the codebook.
The minimum-entropy codebook
is the one with minimum cost plus distortion.

The combinatorial problem of finding this codebook reduces to a
well-studied problem called the fixed-cost $k$-median problem, for which
we propose a new approximation algorithm with a good provable
performance guarantee.  We use this algorithm to find a good codebook.
We compare our partition algorithm, which we will call Greedy k-median
(GKM), with First Fit, First Fit followed by k-means post-processing
steps, and two plausible versions of GKM combined with Lloyd's k-means
algorithm.  By simply substituting GKM for First Fit in the partitioning
step, we reduce the size of the compressed documents in our test suite
by 17\%.

\section{PREVIOUS WORK}

Previous systems\cite{witten:94} used a simple codebook generation algorithm which we
call the First Fit algorithm. This algorithm works as follows: 
\begin{tabbing}
\small
FIRST-FIT(S)\\
1. Let the set of equivalence classes $P\leftarrow \emptyset$ \\
2. {\em for each\/}  \=glyph in the document $c \in S$\\
3. \>{\em \ \ \ do\/} \=find a $p \in P$ such that $distance(p,c) < T $\\
(the distance between p and c is defined as the distance between the
first element of\\ p and c.)\\
4.              \>    \>{\em if \/} there is such a $p$, add $c$
to $p$, {\em continue\/} with next $c$\\
5.  \> \> {\em else\/}  $P\leftarrow P\cup\{\{c\}\}$\\
6.   return $P$\\
\end{tabbing}

The First Fit algorithm runs in $O(mn)$ time ($m$ is the
number of equivalence classes, $n$ is the number of glyphs).
The algorithm is fast, but does not have a good performance guarantee relative
to the optimal partition.  In the worst case, the number of equivalence classes
is $n$ times worse than the optimal partition.
For example, consider the glyph list $L=\{a,b,c,d\}$
and matching pairs $\{a,b\},\{b,c\},\{b,d\}$.
Notice that $a$ does not match directly with $c$ and $d$.
The First Fit algorithm partitions $a$ and $b$ into an equivalence class.
Because $c$ and $d$ do not match to $a$,
which is the first element of the partition $\{a,b\}$,
two partitions are created for $c$ and $d$.
So the final partition is $\{\{a,b\},\{c\},\{d\}\}$.
If we change the order of the list to $\{b,a,c,d\}$,
however, the algorithm finds the partition $\{\{b,a,c,d\}\}$.

Several compression systems, including MGTIC~\cite{witten:src} and
CDIS~\cite{zhang:pe96}, combine a bitmap averaging method with the
First-Fit algorithm to form a multi-pass pattern classification method.
For each equivalence class, all of the bitmaps in the class are averaged
and the result is thresholded to obtain a new pattern.  The glyphs are
then repartitioned, with each glyph being assigned to the first matching
pattern.  Patterns with no matched glyphs are deleted.  This process is
iterated until little or no improvement results.

We also test a variant of this algorithm which reassigns each glyph to
its {\em best\/} matching pattern.  This algorithm is a specialization
of the Generalized Lloyd (k-means) algorithm for vector quantizer
design, because the repartitioned equvalence classes satisfy the
nearest-neighbor condition~\cite{gersho}.  We call this algorithm ``{\em
  modified\/} k-means'' because the averaging and thresholding steps do
not necessarily produce an optimal pattern for each equivalance class.
This can degrade the performance of the algorithm.

\iffalse % minimizing distortion for a given k IS a way to solve the fixed cost
         % k-median problem  -neal
The strict k-means algorithm is optimized for minimizing the average
distortion for a set of inputs. The minimized distortion, however, is
not equal to minimum number of equivalence classes, nor is this equal to
minimum total cost given by the fixed cost k-median problem. So the
k-means algorithm does not provide a solution for our fixed cost
k-median probelm.  The modified k-means First Fit algorithm does not
change the previous worst case analysis of the First Fit algorithm.
\fi

\section{THE GREEDY K-MEDIAN ALGORITHM}
The {\em weighted fixed-cost $k$-median\/} problem is the following.
Let $G=(V,E)$ be a directed or undirected graph 
with non-negative edge weights $d:E\rightarrow R$
and non-negative vertex weights $c:V\rightarrow R$.
For any subset $S$ of vertices,
define the {\em cost\/} of $S$ to be $c(S) = \sum_{w\in S} c(w)$;
define the {\em distortion\/} of $S$ to be $d(S) = \sum_{v\in V} d(v,S)$
where $d(v,S) = \min_{w\in S} d(v,w)$.
The problem is to find a set
minimizing the cost plus the distortion.

Our problem of finding a minimum-entropy codebook
reduces to this problem in a directed graph
with a vertex $v$ for each glyph in the document.
The cost of $v$ is the cost of the glyph's pattern;
the cost of edge $(u,v)$ is the distance between the glyph of $u$
and the pattern of $v$'s glyph.
Any subset of vertices in the graph then represents a codebook
of corresponding cost and distortion.
Because of the properties of glyphs and their patterns,
the weights in this graph satisfy the usual triangle inequality,
as well as $c(v) \le c(u) + d(u,v)$ for all $u,v$.

\subsection{Previous work.}
Hochbaum \cite{hochbaum:82} gives a polynomial-time algorithm
that finds a set for which the cost plus the distortion
is at most $1+\ln n$ times the minimum possible.
Roughly, Hochbaum observes that the problem reduces
to a traditional weighted set-cover problem with exponentially many sets.
She then argues it suffices to consider only quadratically many of these sets
and obtains a cubic-time algorithm by adapting
the weighted greedy set cover algorithm of Chvatal \cite{Chvatal79}.

\newcommand{\opt}{\mbox{\sc OPT}}

For the variant in which the desired cost $k$ is specified,
and the problem is to minimize the resulting distortion,
Lin and Vitter \cite{LV92} give a polynomial-time approximation algorithm
with the following performance guarantee:
given $\epsilon>0$, the algorithm returns a set
of cost at most $(1+1/\epsilon)(1+\ln(|V|)) k$ with distortion at most
$1+\epsilon$ times the minimum distortion for a set of cost $k$.
They present a related algorithm for the fixed-cost problem
that returns a set with cost plus distortion bounded by
$$(1+\epsilon)d(\opt) + (1+1/\epsilon)(1+\ln(|V|)) c(\opt),$$
where $\opt$ is the optimal set.
These algorithms each solve a linear program
and then round the solution using the set-cover algorithm of Chvatal.

\subsection{A New Greedy Algorithm}
Our new algorithm is somewhat simpler to implement than either of the
aforementioned algorithms and provides comparable performance guarantee
in graphs with weights satisfying the triangle inequality.
The algorithm is based on the greedy algorithm for minimizing a
linear function subject to a submodular constraint~\cite{nemhauser}
(a generalization of the greedy set-cover algorithm).

Define the {\em capped distortion} of a set $S$ to be
$$\delta(S) = \sum_{v\in V} \min(d(v,S), c(v)).$$
The algorithm is the following.
\begin{tabbing}
\small
GREEDY-K-MEDIAN($G=(V,E),c,w$)\\
1. $S'\leftarrow \emptyset$ \\
2. {\em do\/} \= $S \leftarrow S'$ \\
3. \> choose $v\in V$ that maximizes 
$(\delta(S)-\delta(S\cup\{v\}))/c(v)$ \\
4.  \> $S'\leftarrow S\cup\{v\}$\\
5.  {\em while} $c(S')+\delta(S') < c(S)+\delta(S)$ \\
6. {\em return} $S$
\end{tabbing}
The loop executes linearly many times and each iteration takes quadratic time.
Thus, GREEDY-K-MEDIAN runs in cubic time.

\subsection{Analysis.}

Fix a graph $G$ with vertex weights $c$ and edge weights $w$.
Recall that $c(S)$ and $d(S)$ denote the cost and distortion, respectively,
of a set $S$.
Let \opt{} denote the set of vertices such that $c(\opt) + d(\opt)$ is minimal.

The main point of the analysis is the following.

\noindent{\bf Claim\/}: {\em
  GREEDY-K-MEDIAN maintains the invariant
  $$\frac{\delta(S)-d(\opt)}{\delta(\emptyset)-d(\opt)}
  \le \exp(-c(S)/c(\opt)).$$
  }

\noindent{\bf Proof\/}: At the beginning of any iteration, adding to $S$ all of
the vertices in \opt{} would result in a set with $\delta(S) = d(\opt)$.
Thus, since $\delta$ is supermodular,
some vertex $w$ in $\opt$ can be added such that
$$\frac{\delta(S)-\delta(S\cup\{w\})}{c(w)}
\ge \frac{\delta(S) - d(\opt)}{c(\opt)}.$$
Together with the choice of $v$ (and a little algebra)
it follows that
$$\delta(S\cup\{v\})-d(\opt) \le (\delta(S)-d(\opt))(1-c(v)/c(\opt)).$$
The result follows inductively, using $1-c(v)/c(\opt) \le \exp(-c(v)/c(\opt))$.
With a little more work (details are in the full paper), one can show:

\noindent{\bf Corollary\/}: {\em
  The set $S$ returned by GREEDY-K-MEDIAN has cost plus distortion bounded by
  $$d(\opt) +
  \left(1+\ln\frac{\delta(\emptyset)}{c(\opt)}\right)\times c(\opt).$$
  }

If the weights satisfy the triangle inequality,
it is not hard to see that
$c(\opt) + d(\opt) \ge \max_v c(v) \ge \delta(\emptyset)/|V|$.
Thus, the previous corollary and a little algebra imply:

\noindent{\bf Corollary\/}: {\em
  Given a graph with edge and vertex weights satisfying the triangle
  inequality, the set $S$ returned by GREEDY-K-MEDIAN has cost plus
  distortion bounded by
  $$2\times d(\opt) + (1+\ln |V|)\times c(\opt).$$
  }

In such graphs,
if the cost of the optimal solution is much smaller than the distortion
(as one might expect in a problem corresponding to a reasonably long document),
these performance guarantees are fairly strong.
In particular, when the ratio between the two is about $\ln |V|$,
the performance guarantee is constant;
when the ratio grows larger, the performance guarantee tends to 1
(this follows from the first corollary).

\subsection{GKM with k-means improvement}
We next combine GKM with the k-means algorithm as follows:
\newcommand{\rate}{\mbox{\rm rate}}
\begin{tabbing}
\small
GKM WITH MODIFIED K-MEANS(G,c,d)\\
0. define $\rate(S,v) = (\delta(S)-\delta(S\cup\{v\}))/c(v)$\\
1. $S'\leftarrow \emptyset$\\
2. {\em do\/} \= $S \leftarrow S'$\\
3. \> choose $v'\in V$ that maximizes $\rate(S,v')$\\
4. \> {\em do\/} \= $v \leftarrow v'$\\
5. \> \> $v' \leftarrow $ the centroid of
$\{w \in V | d(w,v) = \min(d(w,S), c(w))\}$\\
6. \> {\em while} $\rate(S,v') > \rate(S,v) + \epsilon$\\
7. \> $S'\leftarrow S\cup\{v\}$\\
8.  {\em while} $c(S')+\delta(S') < c(S)+\delta(S)$ \\
9. return $S$
\end{tabbing}
The centroid in step 5 is (some estimation of) the {\em optimal} center for the
set of vertices (glyphs) assigned to the vertex (pattern) $v$.  The algorithm
considers it as a possible alternative to $v$.  Similarly, we can use the
modified k-means algorithm as a postprocessing step:

\small
GKM FOLLOWED BY MODIFIED K-MEANS\\
1. Compute the initial equivalence classes by GREEDY-K-MEDIAN.\\
2. Average the glyphs in each equivalence class and creating an ideal
  pattern for each  class.\\
3. Reclassify the glyphs by matching each glyph with its closest
  ideal pattern.    Delete empty classes. \\
4. Go to step 2 while the number of
  equivalence classes is decreased by at least some constant $k$.\\
\normalsize

\begin{table}[tbp]
\small
  \begin{center}
    \leavevmode
  
    \begin{tabular}{lrrrrr}
      Test document&\multicolumn{3}{c}{Codebook size}
       &\multicolumn{2}{c}{Decrease in codebook size} \\ \cline{2-6} 
       &F. F. &K-means&GKM &F.F. -- K-means&F.F.
       -- GKM\\ \hline
      CCITT1&177&157&162&11.3\%&8.5\%\\
      CCITT4&164&163&224&0.6\%&-36.6\%\\
      BROOKS1&408&386&281&5.4\%&31.1\%\\
      BROOKS2&365&339&287&7.1\%&21.4\%\\
      BROOKS3&332&309&247&6.9\%&25.6\%\\
      BROOKS4&371&343&290&7.5\%&21.8\%\\
      LAMBDA1&614&583&313&5.0\%&49.0\%\\
      LAMBDA2&214&180&120&15.9\%&43.9\%\\
      LAMBDA3&713&677&331&5.0\%&53.6\%\\
      LAMBDA4&648&596&368&8.0\%&43.2\%\\ \hline
      Average&&&&7.3\%&26.2\%\\ 
 
    \end{tabular}
    \caption{\em \small
      Codebook size for the First Fit algorithm with and without
      modified k-means optimization and the GKM algorithm. With the
      exception of CCITT4, GKM dominates First Fit.  On average, GKM has
      a 26\% smaller codebook than First Fit, and a 19\% smaller
      codebook  than First Fit with modified k-means.  }
    \label{tab:one_pass_two_pass}
  \end{center}
\end{table}

\section{EXPERIMENTAL RESULTS}
We have implemented the GKM algorithm with and without modified k-means
postprocessing.  $cost(c)$ is estimated as the area of each glyph.  We
compute $cost(c|p)$ using the spatial sampling error based cross entropy
measure mentioned above~\cite{zhang:96}.  We also implemented the First
Fit algorithm with multi-pass optimization as used in the MGTIC
system~\cite{witten:94}.  We have tested these configurations on ten
high quality scanned document pages with over 20,000 glyphs. Two of the
pages, CCITT1 and CCITT4, are from the CCITT test image set, scanned at
200 dpi. The other eight pages, the BROOKS series and the LAMBDA series, are
from the MIT AI lab's technical document collection and are scanned at 300
dpi. We measured the performance of each codebook generation algorithm
according to its use in document image compression. For a fair
comparison, we adjusted pattern matching thresholds in each case so
that no characters were misclassified.

Table 1 compares the sizes of the codebook produced by the First Fit
codebook generation algorithm with and without the multi-pass
optimization, and our GKM algorithm. In each case we used the same underlying
pattern matching algorithm, EPM~\cite{zhang:96}.  We use the first match
version of the modified k-means optimization (used in
MGTIC~\cite{witten:94}) because it produces slightly better results and
runs faster than the best match version.  Compared with the First Fit
algorithm, GKM reduces the codebook size by an average of 26\%, a
remarkable improvement. Notice that for CCITT4, the number of
patterns grows. In this case, GKM shifts the costs of coding one glyph
given another glyph to the costs of patterns while minimizing the
total cost. The modified k-means optimization reduces
the codebook size by an average of 7.2\%, a smaller gain than GKM.

\begin{table}[tbp]
\small
  \begin{center}
    \leavevmode
  
    \begin{tabular}{lrrr}
      Test document&\multicolumn{2}{c}{Codebook size}
       &Decrease in codebook size \\  \cline{2-4}
       &GKM &GKM k-means&GKM--GKM k-means\\ \hline
       CCITT1&162&142&12.3\%\\
       CCITT4&224&138&38.4\%\\
       BROOKS1&281&275&2.1\%\\
       BROOKS2&287&226&21.3\%\\
       BROOKS3&247&194&21.5\%\\
       BROOKS4&290&219&24.5\%\\
       LAMBDA1&313&303&3.2\%\\
       LAMBDA2&120&110&8.3\%\\
       LAMBDA3&331&314&5.1\%\\
       LAMBDA4&368&366&0.5\%\\ \hline
       Average&&&13.7\%\\ 
 
    \end{tabular}
    \caption{\small \em
      Codebook size for GKM and  GKM followed by modified k-means. GKM
      followed by modified k-means results in smaller number
      of pattern for all the test documents. The average
      improvement is  13.7\%.
      }
    \label{tab:naive_g}
  \end{center}
\end{table}

Table 2 compares the sizes of the codebook produced by the plain GKM
algorithm and the GKM followed by modified  k-means algorithm. GKM
followed by  modified  k-means is  better than plain GKM. GKM with k-means (not shown)
performed slightly worse than plain GKM. The mixed results of the
k-means algorithm may be due to thresholding effects.

\begin{table}[tbp]
\small
  \begin{center}
    \leavevmode
  
    \begin{tabular}{lrrrrrr}
      Test
      document&\multicolumn{2}{l}{Lossy}&\multicolumn{2}{l}{Lossless}&\multicolumn{2}{c}{Size of output}\\
      &\multicolumn{2}{l}{compression ratio} &\multicolumn{2}{l}{compression ratio}&GKM/F.F. &GKM/F.F.\\ \cline{2-5}
       &F.F.&GKM&F.F.&GKM&(lossy)&(lossless)\\ \hline
       CCITT1&93.7:1&95.7:1&34.5:1&35.4:1&98.2\%&97.5\%\\
       CCITT4&54.9:1&44.5:1&11.2:1&10.7:1&123.4\%&104.7\%\\
       BROOKS1&87.2:1&105.4:1&42.8:1&47.5:1&82.7\%&90.1\%\\
       BROOKS2&75.2:1&83.8:1&21.8:1&23.1:1&89.7\%&94.4\%\\
       BROOKS3&80.3:1&90.9:1&24.4:1&23.1:1&88.3\%&105.6\%\\
       BROOKS4&72.5:1&80.1:1&20.8:1&22.0:1&90.5\%&94.5\%\\
       LAMBDA1&54.9:1&89.8:1&30.5:1&39.3:1&61.1\%&77.6\%\\
       LAMBDA2&145.0:1&236.7:1&79.4:1&101.9:1&61.6\%&77.9\%\\
       LAMBDA3&48.0:1&68.0:1&13.6:1&20.6:1&70.5\%&66.0\%\\
       LAMBDA4&47.9:1&75.5:1&24.0:1&29.7:1&63.4\%&80.8\%\\ \hline
       Average&&&&&82.9\%&88.9\%\\ 
 
    \end{tabular}
    \caption{\small \em
      Overall compression ratios for plain GKM and First Fit. GKM
      improves compression significantly. The average lossy output of GKM is
      about 83\% of First Fit's output.
      }
    \label{tab:ratio}
  \end{center}
\end{table}

Table 3 compares overall compression ratios achieved using GKM and First
Fit. We used the same lossy glyph position coding
scheme~\cite{zhang:ep96} and Entropy-based Pattern Matching
(EPM)~\cite{zhang:96} to compress the test images. With the exception of
CCITT4, GKM dominated First Fit. The use of GKM reduced the size of the
compressed documents by an average of 17\%.
 
\begin{figure*}[tbp]
\small
  \begin{center}
    \leavevmode   

   {
     %Original\\
     %\framebox{\psfig{figure=1_original.eps}} \\
     Reconstructed image using First Fit algorithm\\
     \framebox{\psfig{figure=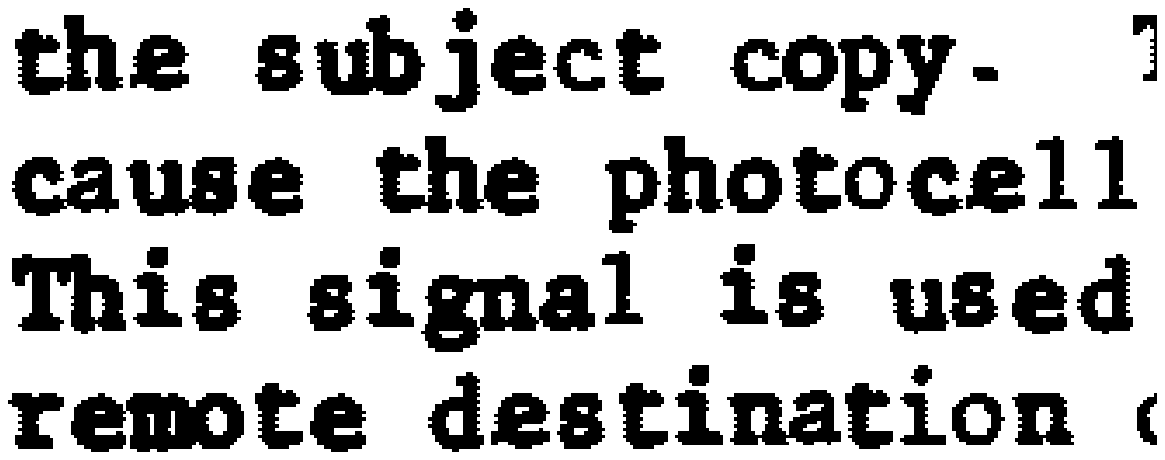,width=3.8in}}\\
     Reconstructed image using GKM\\
    \framebox{\psfig{figure=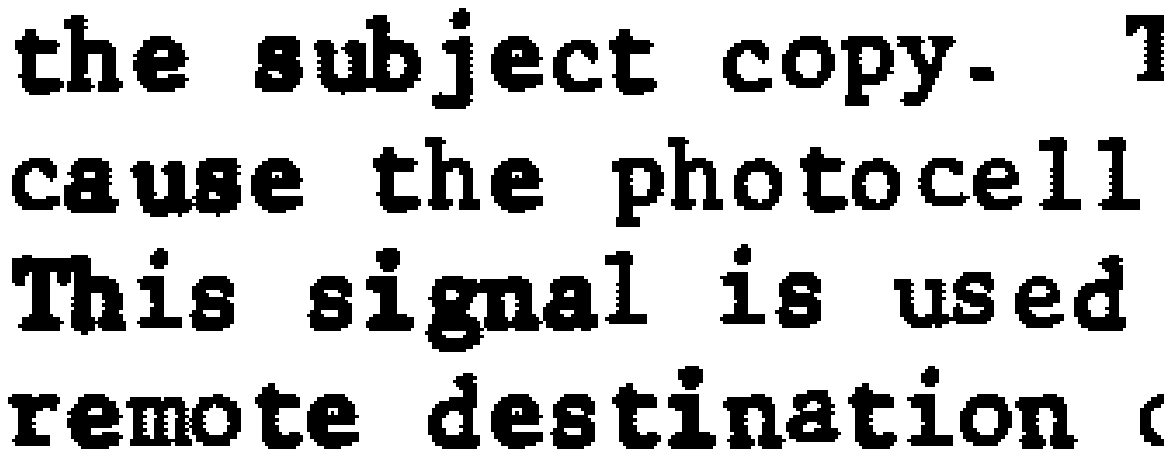,width=3.8in}}
    }
  
   \caption{\small \em A randomly selected zoomed portion of CCITT1 reconstructed
     by First Fit and GKM. It is clear that First Fit and GKM pick
     different glyphs as the representitive patterns to reconstruct the
     original image. In most cases, GKM picks a better quality bitmap.}
    \label{zoom1}
  \end{center}
\end{figure*}

We also compared reconstructed images resulting from plain First Fit and
GKM. As you can see in Figure 1, GKM usually picks a better quality
bitmap as the pattern, resulting in better reconstructed images.

\section{DISCUSSION} 
GKM performed surprising well compared to the popular First Fit
algorithm in our experiments. We had expected to improve compression
by a few percent, but we were unprepared for the 17\% improvement in
overall compression performance that we saw. Clearly the glyph
clustering algorithm is more important than previously realized.

K-means postprocessing steps helped First Fit, but did not help GKM,
presumably because GKM does a good job on its own. We had believed at
first that K-means steps could not degrade a codebook, because
of the performance guarantees associated with this
algorithm. However, after seeing codebooks degraded, we realized that
the distortion measure that K-means is guaranteed not to degrade is
not ours, that our thresholded (non-linear!) distance measures do not
fit into the K-means framework, and that K-means minimizes distortion,
not library size. Use this ``universal'' algorithm with caution.

GKM runs in cubic time, taking several minutes to partition a
single dense page of text on a moderately aged DEC Alpha
workstation. This is far too long, although since only encoding is
slow, this extra time could be tolerable for some applications. We
expect investigations into efficient data structures for GKM to be
fruitful.

A shortcoming in our model that we would like to address (at a later
date) is the fact that we assume that all patterns are equally likely:
there is no explicit pattern indexing cost in the minimization. If we
could include estimates of these costs in the minimization, then
patterns which are likely in some context would be favored over
patterns which are unlikely in some context. Since text entropies are
small compared to glyph entropies, this would be a small effect, but we
might see these costs breaking ties between two patterns which match a
glyph almost equally well, as often occurs with ``l'' and ``1'' and
occasionally ``a'' and ``s''.  In essence we would like to develop a
minimum entropy model of the document which encourages consistent
spelling by taking into account the syntactic structure of the
document.

\end{document}